\documentstyle[11pt,newpaspman,twoside,epsfig]{article}
\def\edcomment#1{\iffalse\marginpar{\raggedright\sl#1\/}\else\relax\fi}
\reversemarginpar

\begin{document}
\title{Optical and near-infrared observations of Anomalous X-ray Pulsars}
 \author{Ferdi Hulleman}
\affil{Utrecht University, P.O. Box 80000, 3508 TA Utrecht, The Netherlands}

\begin{abstract}
We present recent results of our program of deep optical and
infrared observations of AXPs. We find that the counterpart
to AXP 4U~0142+61 has a peculiar spectral energy distribution. The counterpart is remarkably bright in the near-infrared, but is not detected in the B band. 
We also present the possible detection of a second AXP counterpart,
namely that to 1E~2259+586. 
\end{abstract}

\section{Introduction}
Anomalous X-ray Pulsars (AXP) are mysterious objects. We do not
understand the energy source that is responsible for their X-ray emission;
unlike young Crab-like pulsars their spin-down energy is insufficient
and unlike binary X-ray pulsars there is no sign of binary companions
from which they could accrete. In the two models most often considered
AXP are either isolated neutron stars 
accreting from a fossil disk, formed out of supernova debris
(e.g. Chatterjee, Hernquist \& Narayan 2000) or after a common envelope
phase (van Paradijs, Taam \& van den Heuvel 1995), or magnetars, neutron stars
with an ultrahigh magnetic field ($B > 10^{14}$\,Gauss, Thompson \&
Duncan 1996).   

Keck data of 4U~0142+61, the X-ray brightest AXP, revealed a faint
$R=24.99\pm0.07$ source within the 3.9 arcsec radius {\em Einstein}
error circle\footnote{(C. Kouveliotou (priv. comm.) informed us that
the source is also within the smaller {\em Chandra} error circle. See
also the contribution by Juett et~al. in these proceedings.},
that has peculiar optical colours, $V-R=0.63\pm0.11$,
$R-I=1.15\pm0.09$ (Hulleman et~al. 2000). By comparing Keck images
from 1994 and 1999 we found its brightness to be constant to within
0.2 mag ($2\sigma$) in R and its proper motion to be less than 0.03
arcsec per year (again $2\sigma$).    

\section{Recent results}

\begin{figure}[t]
\epsfig{file=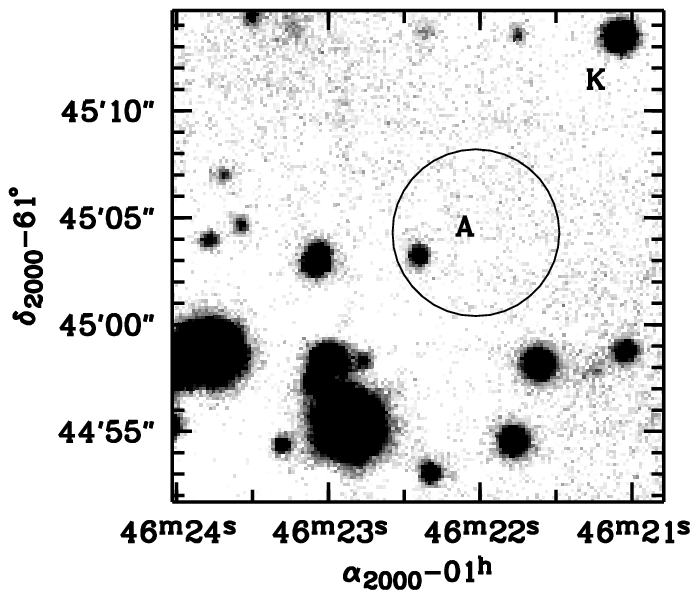,width=2.5in,clip=t}\epsfig{file=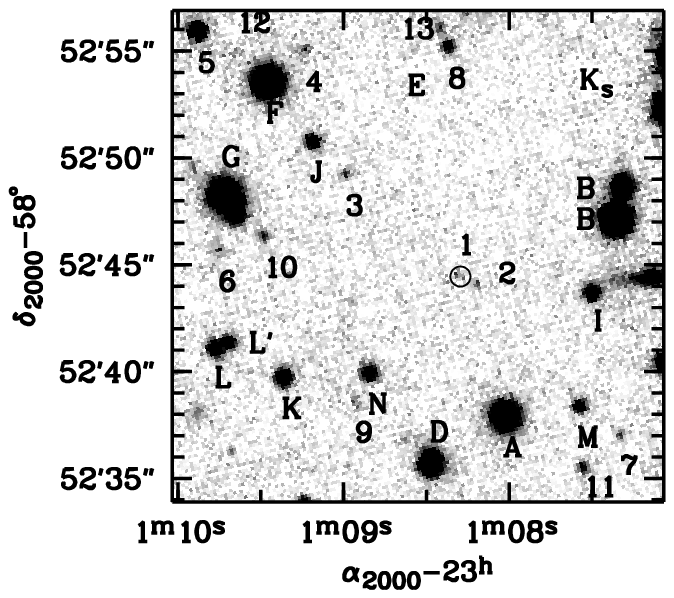,width=2.5in,clip=t}
\caption{\\
Left: Near-infrared image of the field around AXP
4U~0142+61. Overplotted is the {\em Einstein} error circle
(White~et~al. 1987). Star A is the counterpart.\\ 
Right: Near-infrared image of the field around AXP
1E~2259+586. The circle represents the {\em Chandra} position. See
Hulleman et~al. (2001) for details.}
\end{figure}

\subsection{The spectral energy distribution of AXP 4U~0142+61}

We have extended our data set of 4U~0142+61 with photometry in the optical
B band and the near-infrared K band. The source is brighter
than expected in K ($K=19.6$, Fig.~1), but it is not
detected in the B band. None of the currently available models
can explain the observed optical emission (Fig.~2), although we
note that no model for the optical emission from a magnetar is
currently available.

\begin{figure}
\epsfig{file=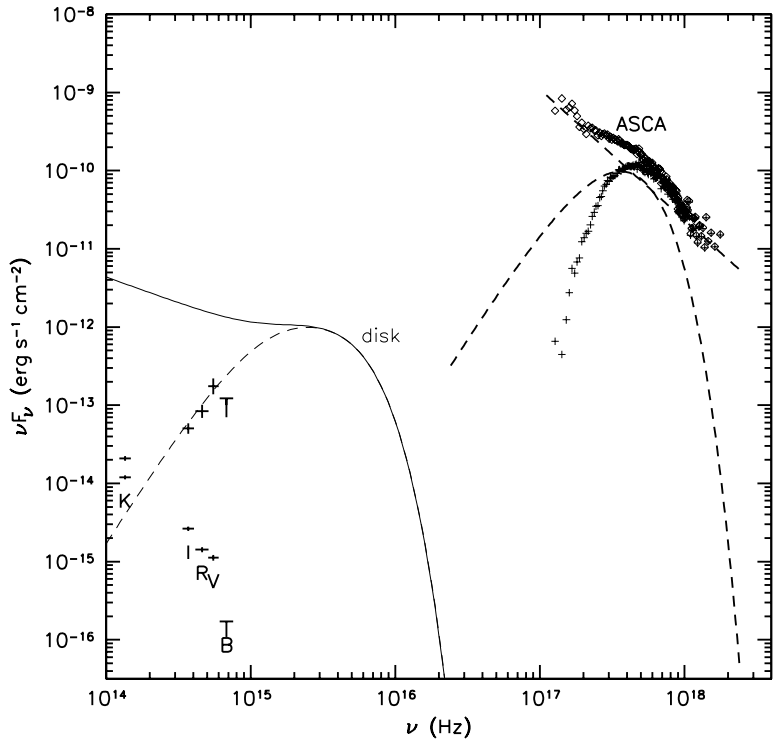,width=4in}
\caption{Observed and extinction corrected
fluxes from AXP 4U~0142+61. Also shown are the unabsorbed blackbody
and power-law components of the model that best fits the X-ray
spectrum. Overplotted are models of the optical emission of a large
disk around an isolated neutron star (solid line) and of a small disk
in a compact binary (thin dashed line).\label{fig1}}
\end{figure}

\subsection{A possible second AXP counterpart.}

Recently a subarcsecond position has been derived for AXP 1E~2259+586
using {\em Chandra} data (Hulleman et~al. 2001). Within the error circle we
find a faint near-infrared source ($K_{\rm s}=21.7\pm0.2$). It is not
detected in the J, I and R bands. We set limits of 23.8, 25.6 and
26.4\,mag in J, I and R respectively. Finally, we note that the near-infrared
to \mbox{X-ray} flux ratio is similar for both AXPs.

\acknowledgements
I would like to thank Marten van Kerkwijk and Shri Kulkarni for useful
comments.



\end{document}